\documentclass[aps,prl,twocolumn,superscriptaddress,floatfix]{revtex4-1}

\usepackage[hidelinks]{hyperref}

\usepackage{graphicx}
\usepackage{dcolumn}
\usepackage{bm}
\usepackage{siunitx}
\usepackage[utf8]{inputenc}
\usepackage[T1]{fontenc}
\usepackage{mathptmx}
\usepackage{etoolbox}
\usepackage{float}
\usepackage{booktabs}
\usepackage{multirow}
\usepackage{color}
\usepackage{soul}
\usepackage{ulem}

\usepackage{bm}
\usepackage{xcolor}
\usepackage{verbatim}
\usepackage{physics}
\usepackage{appendix}
\usepackage{soul} 
\usepackage{chemformula}
\usepackage{siunitx}
\usepackage{amsmath}

\usepackage[thinc]{esdiff}
\usepackage{todonotes}
\usepackage{silence}
\usepackage{array}

\newcolumntype{P}[1]{>{\centering\arraybackslash}p{#1}}

\DeclareSIUnit\angstrom{\text {Å}}

\newcommand{\beq}{\begin{equation}}
\newcommand{\eeq}{\end{equation}}
\newcommand{\beqs}{\begin{equation*}}
\newcommand{\eeqs}{\end{equation*}}
\newcommand{\beqa}{\begin{eqnarray}}
\newcommand{\eeqa}{\end{eqnarray}}
\newcommand{\beqas}{\begin{eqnarray*}}
\newcommand{\eeqas}{\end{eqnarray*}}
\def\bals#1\eals{\begin{align*}#1\end{align*}}
\def\bal#1\eal{\begin{align}#1\end{align}}

\makeatletter
\def\@email#1#2{%
 \endgroup
 \patchcmd{\titleblock@produce}
  {\frontmatter@RRAPformat}
  {\frontmatter@RRAPformat{\produce@RRAP{*#1\href{mailto:#2}{#2}}}\frontmatter@RRAPformat}
  {}{}
}%
\makeatother
\begin{document}

\title{First-Principles Nanocapacitor Simulations of the Optical Dielectric Constant in Water Ice}
\author{Anthony Mannino}
\affiliation{Physics and Astronomy Department, Stony Brook University. Stony Brook, New York 11794-3800, USA}
\affiliation{Institute for Advanced Computational Science, Stony Brook, New York 11794-3800, USA}

\author{Graciele M. Arvelos}
\affiliation{Instituto de F\'isica Te\'orica, Universidade Estadual Paulista (UNESP), S\~ao Paulo, SP 01140-070, Brazil}
\author{Kedarsh Kaushik}
\affiliation{Physics and Astronomy Department, Stony Brook University. Stony Brook, New York 11794-3800, USA}
\affiliation{Institute for Advanced Computational Science, Stony Brook, New York 11794-3800, USA}
\author{Emilio Artacho}
\affiliation{CIC Nanogune BRTA and DIPC, Tolosa Hiribidea 76, 
             20018 San Sebastian Spain}
\affiliation{Ikerbasque, Basque Foundation for Science, 
48011 Bilbao, Spain}
\affiliation{Theory of Condensed Matter,
             Cavendish Laboratory, University of Cambridge, 
             J. J. Thomson Ave, Cambridge CB3 0HE, United Kingdom}
\author{Pablo Ordejon}
\affiliation{\textit{Catalan Institute of Nanoscience and Nanotechnology - ICN2 (CSIC and BIST), Campus UAB, Bellaterra, E-08193, Barcelona, Spain}}
\author{Alexandre R. Rocha}
\affiliation{Instituto de F\'isica Te\'orica, Universidade Estadual Paulista (UNESP), S\~ao Paulo, SP 01140-070, Brazil}
\author{Luana S. Pedroza}
\affiliation{Instituto de F\'isica, Universidade de S\~ao Paulo, SP 05508-090, Brazil}
\author{Marivi Fernández-Serra}
\affiliation{Physics and Astronomy Department, Stony Brook University. Stony Brook, New York 11794-3800, USA}
\affiliation{Institute for Advanced Computational Science, Stony Brook, New York 11794-3800, USA}
\email{marivi.fernandez-serra@stonybrook.edu}
\email{luana@if.usp.br}

\date{\today}

\begin{abstract}
We introduce a combined density functional theory (DFT) and non-equilibrium Green’s function (NEGF) framework to compute the capacitance of  nanocapacitors and directly extract the dielectric response of a sub-nanometer dielectric under bias.
We identify that at the nanoscale conventional capacitance evaluations based on stored charge per unit voltage suffer from an ill-posed partitioning of electrode and dielectric charge.
This partitioning directly impacts the geometric definition of capacitance through the capacitor width, which in turn makes the evaluation of dielectric response uncertain.
This ambiguous separation further induces spurious interfacial polarizability when analyzed via maximally localized Wannier functions. 
Focusing on crystalline ice, we develop a robust charge-separation protocol that yields unique capacitance-derived polarizability and dielectric constants, unequivocally demonstrating that confinement neither alters ice’s intrinsic electronic response nor its insensitivity to proton order.
Our results lay the groundwork for rigorous interpretation of capacitor measurements in low-dimensional dielectric materials.

\end{abstract}

\maketitle

The dielectric response of nanoconfined water has been the subject of a large number of recent studies \cite{Bresme2021,Michaelides2024,Deissenbeck2023,Galli2013,Zubeltzu2016,DeLuca2016,Schlaich2016,Zhang2018,Motevaselian-ACS-Nano-2020}.
These works were mainly motivated by experimental results measuring the out-of-plane dielectric constant of water between graphene and BN capacitor plates \cite{fumagalli2018}. 
This work concluded that the dielectric constant of water thin films
decreases with thickness from its bulk value of $\varepsilon=80$ to a strikingly small $\varepsilon=2.1$ for films less than 15 {\AA} in width.
Subsequent computational works have tried to understand why the polarization response of water to an external electric field in the nano- or sub-nano- confinement range is
so small.
All of these studies relied on the determination of the dielectric constant through changes in the polarization of the thin film under an applied external field \cite{Zubeltzu2016,Zubeltzu25} or simulations under a constant electric displacement \cite{Zhang2018,Michaelides2024}.
Different explanations have been proposed, ranging from the structural properties of the confined films \cite{Michaelides2024} to the cancellation of anisotropic long-range dipole correlations near the confined film surfaces \cite{olivieri-JPCL-2021}.
Very recently however, Zubeltzu {\it et al.} \cite{Zubeltzu25}, 
highlighted that all of these works rely on making predictions that depend on ill-defined properties when taken to the sub-nanoscale.
At the macroscopic level, determining the dielectric constant of the material filling a capacitor is straightforward: one measures the stored charge \(Q\) on two capacitor plates of area \(A\) separated by a well‐defined width \(w_0\) when an external bias \(V\) is applied, and uses
\begin{equation}
  \varepsilon_{\perp} \;=\; \frac{C}{\,C_{0}\!}
  \;=\;\frac{Q/V}{\,\varepsilon_{0}A/w_0\,}\;.
  \label{eq:eps_macro}
\end{equation}
Here, $C$ and $C_0$ are the capacitances of the full and empty capacitor, $\varepsilon_0$ is the permitivity of free space, \(w_0\) is the plate separation, and \(Q\) lives entirely on the metallic surfaces.
%
%
However, when we shrink the device to the nanometer or subnanometer scale, $w_0$ and $Q$, the two key ingredients of that simple macroscopic picture become ill‐defined.
Here, different definitions of $w_0$ can vary by several \AA\, leading to a \(20\%\) (or larger) uncertainty in \(C_{0}=\varepsilon_{0}A/w_0\) even before any dielectric is added (see Supplementary Material SM).
In addition, the metal electron density spills out into the first molecular layers of the dielectric, and the polarization charge of the dielectric overlaps spatially with the metal’s induced charge.
Any attempt to split the total electron-density change into electrode-charge and dielectric-charge becomes ambiguous.
Because of these two intertwined ambiguities —effective separation and charge partition— applying Eq.~\ref{eq:eps_macro} to sub‐nanometer capacitors can produce spurious results. 

In this context, Zubeltzu {\it et al.} \cite{Zubeltzu25} following what was done for 2D materials \cite{Tian2020}, proposed that the true measure of dielectric response for a nanometer‐thick film is the two‐dimensional (2D) polarizability,
\begin{equation}
  \alpha_{\perp} \;=\; \frac{\partial \mathcal{P}_{2D}}{\partial D_{\perp}},
  \label{eq:alpha_def_intro}
\end{equation}
where \(\mathcal{P}_{2D}\) is the dipole moment per unit area and \(D_{\perp}\) is the applied displacement field.  One can relate \(\alpha_{\perp}\) to an intrinsic \(\varepsilon_{\perp}\) of the film via
\begin{equation}
  \varepsilon_{\perp} \;=\;\Bigl(1 - \tfrac{\alpha_{\perp}}{w_{f}}\Bigr)^{-1},
  \label{eq:eps}
\end{equation}
but here \(w_{f}\) —the true thickness of the dielectric film— must itself be consistently defined  at the atomic scale.

All of the above motivates a fully first‐principles approach in which (i) the metal electrodes’ electronic degrees of freedom are treated explicitly, (ii) the metal slabs can sustain different chemical potentials under bias, and (iii) no {\it a priori} film thickness or charge partition is assumed.  
While a subset of these components have been incorporated in some simulation frameworks \cite{ye2021probing,zhu2025dielectric,li2025electron}, only recently it was shown that such computational construction can be realized with a combination of non-equilibrium Green's functions (NEGFs) to
describe the open system and density functional theory to characterize the Hamiltonian of the device \cite{pedroza2018bias,arvelos2024probing,PhysRevBsmeagol,brandbyge2002density}.
In this work, we show that using this methodology, it is possible to accurately obtain the electronic response of a nano-confined dielectric to an applied external bias, hence directly reproducing experimental measurements.
This framework allows us to compute, under two different applied biases, \(\Delta V\), the spatially resolved electron‐density change \(\Delta\rho(\mathbf{r})\) throughout both electrodes and dielectric.  
From \(\Delta\rho(\mathbf{r})\) one can identify the exact centroids of induced charge on each electrode—thus unambiguously defining the plate separation, and  isolate the polarization charge inside the dielectric by comparing with a standalone‐slab reference. 
In doing so, we remove any ambiguity in applying Eqs.~\ref{eq:eps_macro} or~\ref{eq:eps} at the \(\sim\) \AA\ scale.
While the methodology is general and can be applied to any type of dielectric system, we focus in this work on understanding the purely electronic or optical response, $\varepsilon_{\infty}$, of hexagonal ice \textit{Ih} and \textit{XI} \cite{RMPIce,Pamuk2015}.
Note that throughout this work we use $\varepsilon_{\perp}$ to refer to $\varepsilon_{\perp,\infty}$.
We will show that, for the electronic response, ice already reflects most of the physics that should already be present in nanoconfined water capacitors, but makes the overall analysis more simple.

\begin{figure}[!]
    \centering
     \includegraphics[width=1\linewidth]{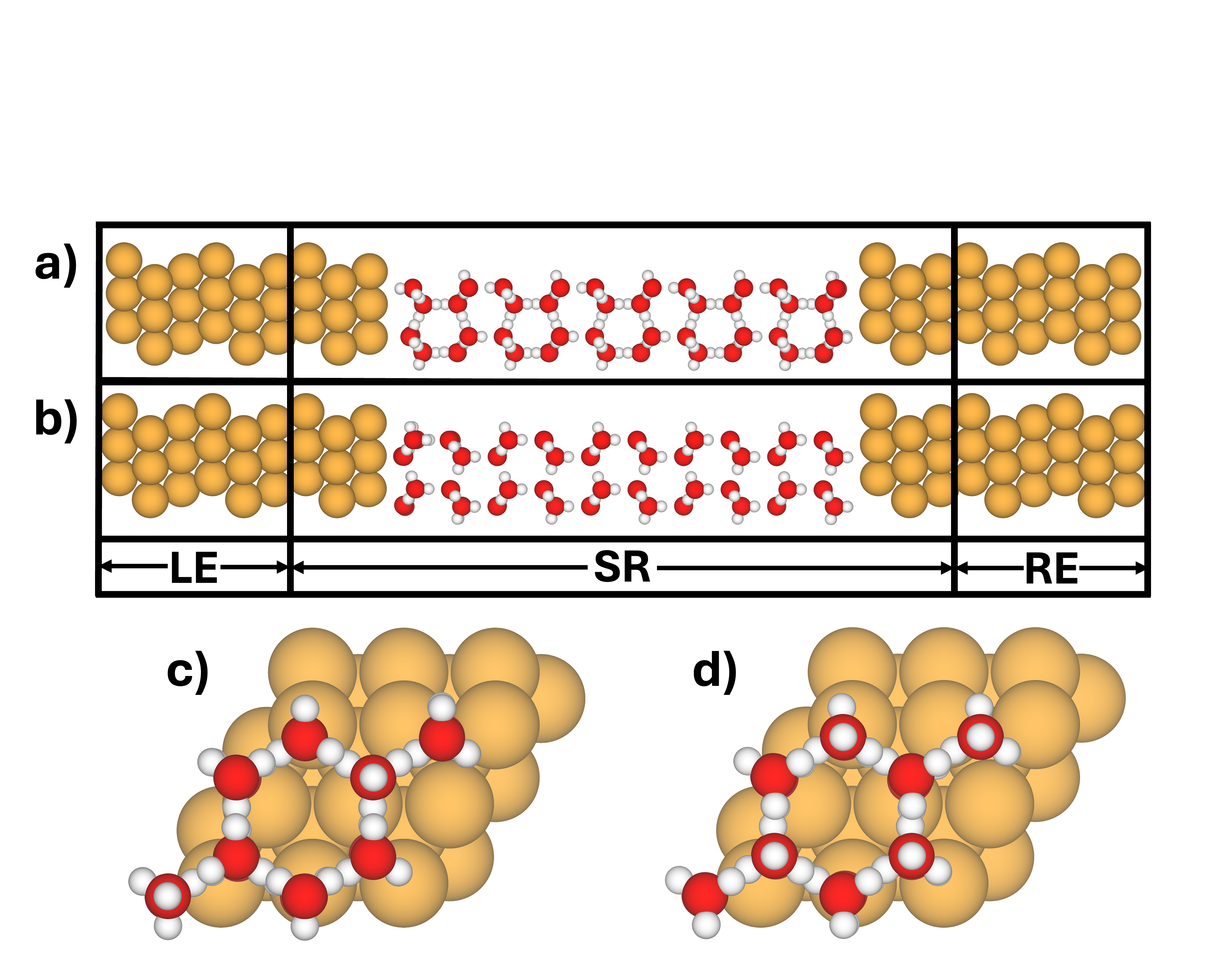}
    \caption{Illustrations of the $\mathbf{(a, c)}$ ice $\it{Ih}$ - Au[111] capacitor and $\mathbf{(b, d)}$ ice $\it{XI}$ - Au[111] capacitor. $\mathbf{(a, b)}$ are side views and $\mathbf{(c, d)}$ are top views. The left and right electrodes (LE and RE, respectively) and the scattering region (SR) are shown.}
    \label{fig:Ice_Structs}
\end{figure}

The simulated plate capacitors consisted of two semi-infinite Au slabs grown along the [111] direction acting as electrodes, and an ice slab \cite{Pamuk2012, Pamuk2015} in between, with its $c$ axis perpendicular to the Au electrode surfaces, as illustrated in Fig. \ref{fig:Ice_Structs}. 
As described in \cite{pedroza2018bias, arvelos2024probing}, the system is split into three regions \cite{caroli}, namely two semi-infinite electrodes, capable of sustaining a different chemical potential if a bias is applied, and a scattering region (SR), which consists of the ice slab and three layers of Au on either side. 
We consider two different proton order configurations for the ice slab: (i) a proton ordered structure with four molecules per unit cell with a net dipole moment along the $c$ axis, that we coin ice \textit{XI} as in \cite{Pamuk2012}. The ice $\it{XI}$ slab consists of 2x2x5 repetitions of this unit; (ii) a zero-dipole moment structure, twice the size of the ice $\it{XI}$ unit cell along the two in-plane directions, with also 5 unit repetitions along the $c$ direction. 
We refer to this structure as the ice $\it{Ih}$ slab, since it mimics the net zero dipole of proton disordered systems.
Both structures are illustrated in Fig. \ref{fig:Ice_Structs}.
These ice slabs can be described as a stack of ten identical ice bilayers.
All calculations were done using the Siesta \cite{soler2002siesta,garcia2020siesta} code which incorporates the Transiesta \cite{brandbyge2002density} method for NEGF calculations. 
We evaluate how the presented results depend on structural relaxations with and without bias in the Supplementary Material (SM).
Additional computational details are also provided in the SM.

For an empty Au‐plate capacitor, we compute capacitance by applying a small bias difference \(\Delta V\) and integrating the induced charge in each electrode:
\begin{equation}
  \Delta Q_{0} \;=\; \int_{z_{\rm in}}^{z_{\rm out}} \Delta\rho_{\Delta V}(z)\,dz,
  \qquad
  C_{0} \;=\; \frac{\Delta Q_{0}}{\Delta V}.
\end{equation}
Here the integration limits \(z_{\rm in}\), \(z_{\rm out}\) are chosen where \(\Delta\rho_{\Delta V}(z)\) vanishes in the bulk electrode (\(z_{\rm in}\)) and inside the gap region (\(z_{\rm out}\)).
The corresponding plate separation \(w_{0}\) is taken as the distance between the two centroids of \(\Delta\rho_{\Delta V}(z)\). 
When ice fills the gap, the total induced charge includes both electrode‐surface charge and polarization charge in the ice. 
To isolate the electrode contribution, we first determine the uniform displacement field change \(\Delta D_{\perp}\) using a half‐full capacitor by fitting the slope of the Hartree potential in the vacuum region, see Fig.~\ref{fig:Ice_Structs} and SM.  
We then apply that same \(\Delta D_{\perp}\) to an isolated ice slab to compute its polarization charge \(\Delta\rho_{\!D}^{\rm slab}(z)\).  Subtracting \(\Delta\rho_{\!D}^{\rm slab}(z)\) from the half-full capacitor \(\Delta\rho_{\Delta V}(z)\) leaves only the electrode‐induced charge. This procedure assumes linearity under bias, which is demonstrated in the SM.
It is important to note that when doing this, we assume that the polarization charge is unchanged by the presence of the metal, which is true for our ice systems as we will show.  
Integrating this difference yields the true electrode charge difference  \(\Delta Q\).
Comparing the induced charge \(\Delta Q\) of the full capacitor with that of the the empty‐capacitor \(\Delta Q_0\) fixes \(\varepsilon_{\perp}^{\rm eff}\), the effective dielectric response of the Au–ice–Au stack via
\begin{equation}
  \varepsilon_{\perp}^{\rm eff}
  \;=\;
  \frac{C\,w}{\,C_{0}\,w_{0}\,}
  \;=\;
  \frac{\Delta Q\,w}{\,\Delta Q_{0}\,w_{0}}\
  \label{eq:eps_eff}
\end{equation}
where \(w\) is defined from the charge‐centroid separation of the electrode-induced charges.
As we will show later, $w_0$ and $w$ do not necessarily need to be the same and in general they are not, even if the inter-plate distance is kept constant.

To isolate the ice slab’s intrinsic 2D polarizability, we use the relation \cite{Zubeltzu25}
\begin{equation}
  \alpha_{\perp} 
  \;=\; 
  (\,w - w_{0}\,) \;+\; \varepsilon_{0}\,A\,\bigl(C_{0}^{-1} - C^{-1}\bigr).
  \label{eq:alpha}
\end{equation} 
$\alpha_{\perp}$ has units of length, and it is an extensive quantity, i.e. depends on the thickness of the measured material.

With the equations above we can compute
$\alpha_{\perp}$ and $\varepsilon_{\perp}^{\rm eff}$
using capacitance calculations from our biased Au electrode-ice devices.
As our dielectric film is crystalline, it is also possible to exactly compute its intrinsic dielectric response $\varepsilon_{\perp}$ using Eq.~\ref{eq:eps}, because here $w_f=c_{\perp}$, the lattice parameter of the system along the normal direction.

We can also compute a spatially resolved $\varepsilon_{\perp}(z)$ directly from the electron density using the polarization analogue to Gauss's Law.

\beq
\label{eq:Gauss}
\div \Delta\mathbf{P} = -\Delta\rho(\mathbf{r})
\eeq
where $\mathbf{P}$ is the 3D polarization vector.
Given the symmetry of the parallel capacitor geometry, we can simplify this to the following form

\beq
\label{eq:P_from_e}
\Delta P (z) = -\frac{1}{A} \int_{z_0}^z \Delta \bar{\rho}(z')dz'
\eeq

\noindent which can then be used to calculate the out of plane dielectric constant as:

\beq
\label{eq:ep_from P}
\varepsilon_{\perp}(z)=\bigg(1-\frac{\Delta P(z)}{\Delta D}\bigg)^{-1}.
\eeq
The constant $z_0$ is determined such that $\varepsilon_{\perp}$ diverges in the electrode. 
$\bar{\rho}$ is the running average of the electronic density over $c_{\perp}$. 
This continuous $\varepsilon_{\perp}(z)$ mixes the dielectric constants from the metal and the ice at the interface when using \(\Delta\rho_{\Delta V}(z)\).
%
%
Using the charge partition scheme previously presented 
it is possible to separate it into electrode and dielectric components.

We can alternatively calculate the local dipole moments of individual molecules using Wannier functions \cite{mostofi2014updated}, which in principle allow us to decouple the dielectric and metal electronic degrees of freedom.
%
%
Then one can use the  Wannier charge centers (WCCs) to compute  
$\mathcal{P}_{2D}$ for each bilayer (BL) and this can be used to compute $\alpha_{\perp}$ as:

\beq
\label{eq:alpha_Wannier}
\alpha_{\perp}^{BL} = \frac{\Delta \mathcal{P}_{2D}^{BL}}{\Delta D}
\eeq
These can then be used to compute a layer-resolved $\varepsilon_{\perp}$ using Eq.~\ref{eq:eps}.

\begin{figure}[!]
    \centering
     \includegraphics[width=1\linewidth]{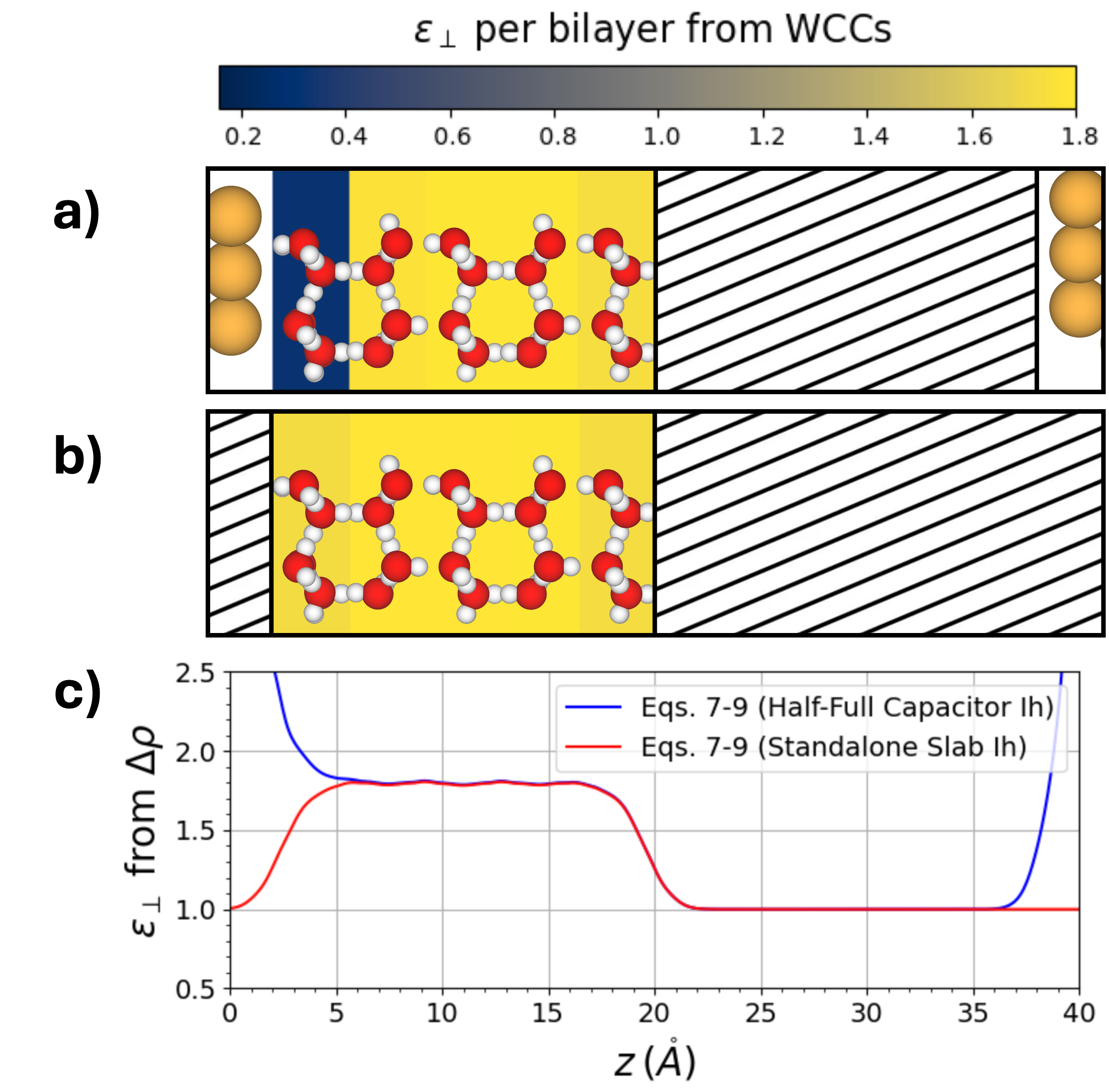}
\caption{Optical dielectric constant $\varepsilon_{\perp}$ in \textbf{(a)} half-full capacitor and \textbf{(b)} standalone slab geometries calculated using two methods. \textbf{(a,b)} The Wannier charge center (WCC) method described by Eqs. \ref{eq:eps} and \ref{eq:alpha_Wannier}. \textbf{(c)} The electron density method described by Eqs. \ref{eq:Gauss}-\ref{eq:ep_from P}. Surface Au electrode planes are at $z = 0$ and $z = 40$ \AA . In both cases finite differences are computed with $\Delta V=1V$.}
    \label{fig:Eps}
\end{figure}

The computed values of the optical dielectric constant for the half-full ice \textit{Ih} - Au capacitor using the electronic density and WCCs methods are shown in Fig. \ref{fig:Eps}.
We also computed the optical dielectric constant using WCCs for an identical ice \textit{Ih} slab placed in vacuum under an applied displacement field identical to that of the capacitor simulation. 
Results for this ice-only slab calculation are also presented in Fig. \ref{fig:Eps}.
In the bulk ice region, all of these methods predict $\varepsilon_{\perp}\approx1.8$, closely matching the known optical dielectric constant of ice and water \cite{petrenko1999physics, Hill1963}.
This is expected, given that the electronic molecular polarizabilities of ice and water are almost identical \cite{DeyuLu}.
This is also the reason why the computed optical dielectric constant is independent of the structural relaxation of the system under bias as shown in the SM.

\begin{figure}[!]
    \centering
     \includegraphics[width=1\linewidth]{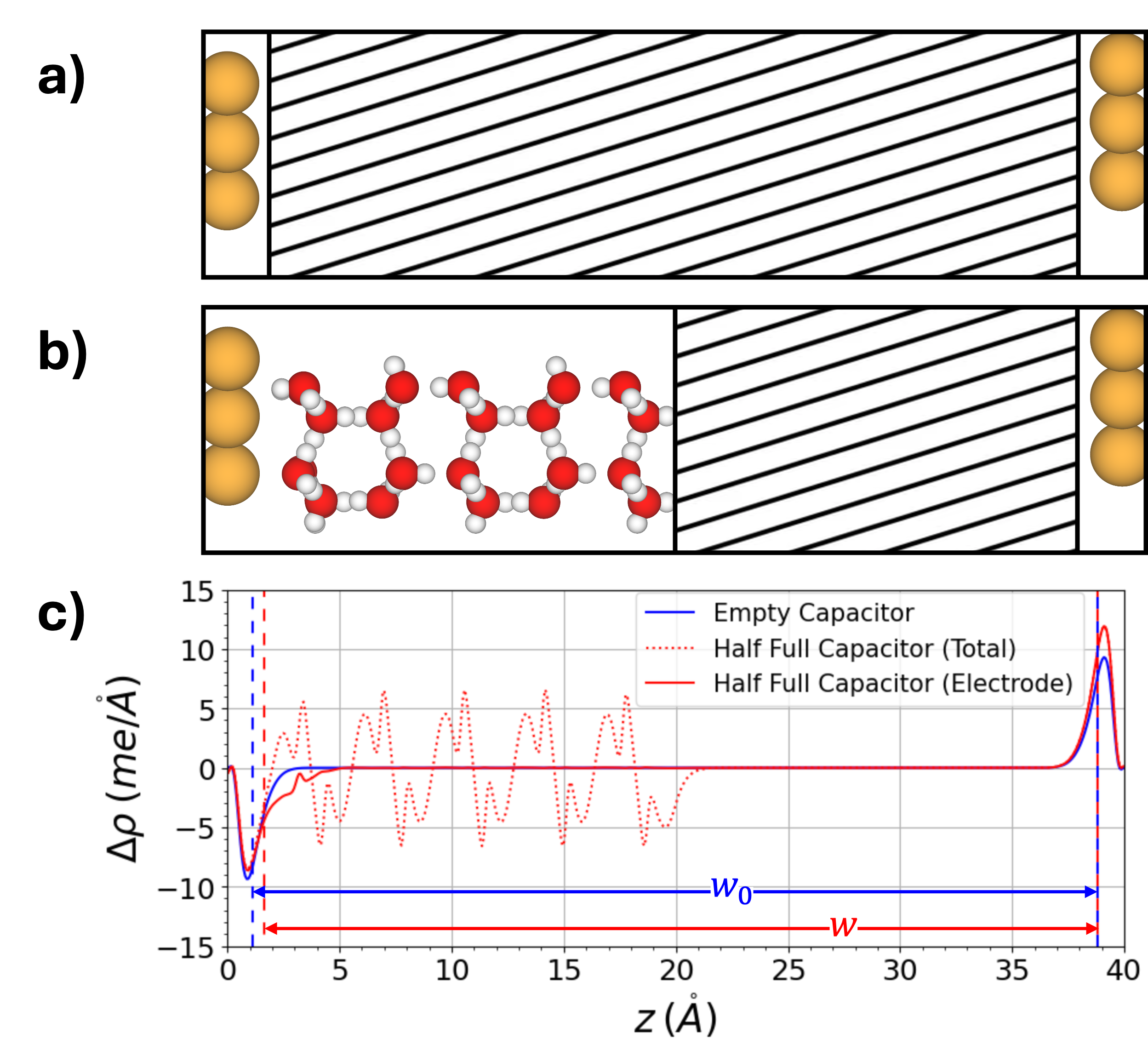}
    \caption{Illustrations of \textbf{(a)} the empty capacitor and \textbf{(b)} the half-full capacitor geometries. \textbf{(c)} Electrode charge density difference for the empty capacitor (solid blue) and the half-full capacitor (solid red) geometries. The total charge density difference for the half-full capacitor geometry is also shown (dotted red). Surface Au electrode planes are at $z = 0$ and $z = 40$ \AA. The vertical dashed lines give the boundaries of the dielectric region, and the capacitor widths, $w$ and $w_0$, for each system are labeled.}
    \label{fig:Q_Cap}
\end{figure}

The WCC results for the half-full capacitor shown in this figure are surprising near the metal/ice interface. 
They suggest that $\varepsilon_{\perp}<1$ ($\alpha_{\perp}<0$) near the interface, indicating an anti-screening response. 
This is at odds with the results for the ice slab, which only see a slight decrease in the optical dielectric constant at the interfaces. 
This discrepancy quickly dissipates away from the interface, with results being insensitive to the method by the third bilayer from the metal interface.
According to these results, ice has a non-homogeneous layer-by-layer polarizability when
in contact with a metal.
If this was the case, subtracting  $\Delta\rho(z)$ for the standalone ice slab from $\Delta\rho(z)$ for the half-full capacitor would
result in large density oscillations near the metal water interface.
This is because the amplitude of the density oscillations corresponds to the difference in the dielectric response.
Note that the standalone ice slab, which is polarized with the same displacement field as the ice slab within the capacitor, results in a uniform layer-by layer polarization.
Results for this subtraction are plotted in Fig. \ref{fig:Q_Cap}.
The remaining charge density difference, while not exempt of small perturbations, closely resembles the characteristics of $\Delta\rho(z)$ for the empty capacitor, which is also plotted in the figure. 
If the optical dielectric response near the metallic interface were much different from that of the slab, as suggested by the WCC results, we would expect to see large oscillations in the dielectric region. 
However, in this case the oscillations are virtually nonexistent, showing that the dielectric response of the ice within the capacitor is virtually the same as the standalone ice slab.
This confirms our initial assumption that the polarization charge is not affected by the presence of the metal.
The most noticeable change is that the electrode charge in contact with ice seems to penetrate deeper into the dielectric region than in the empty capacitor.
These two charge densities can be used to compute $w-w_0 = -0.48 {\AA}$, which is shown in Fig.~\ref{fig:Q_Cap}.
As mentioned earlier atomic relaxations do not modify the value of the optical dielectric constant of ice. 
However, they increase the electronic charge overlap at the ice-Au interface, worsening the error in the surface layer polarizabilities computed using the WCCs.
A detailed analysis of this is presented in the SM.

\begin{figure}[!]
    \centering
     \includegraphics[width=1\linewidth]{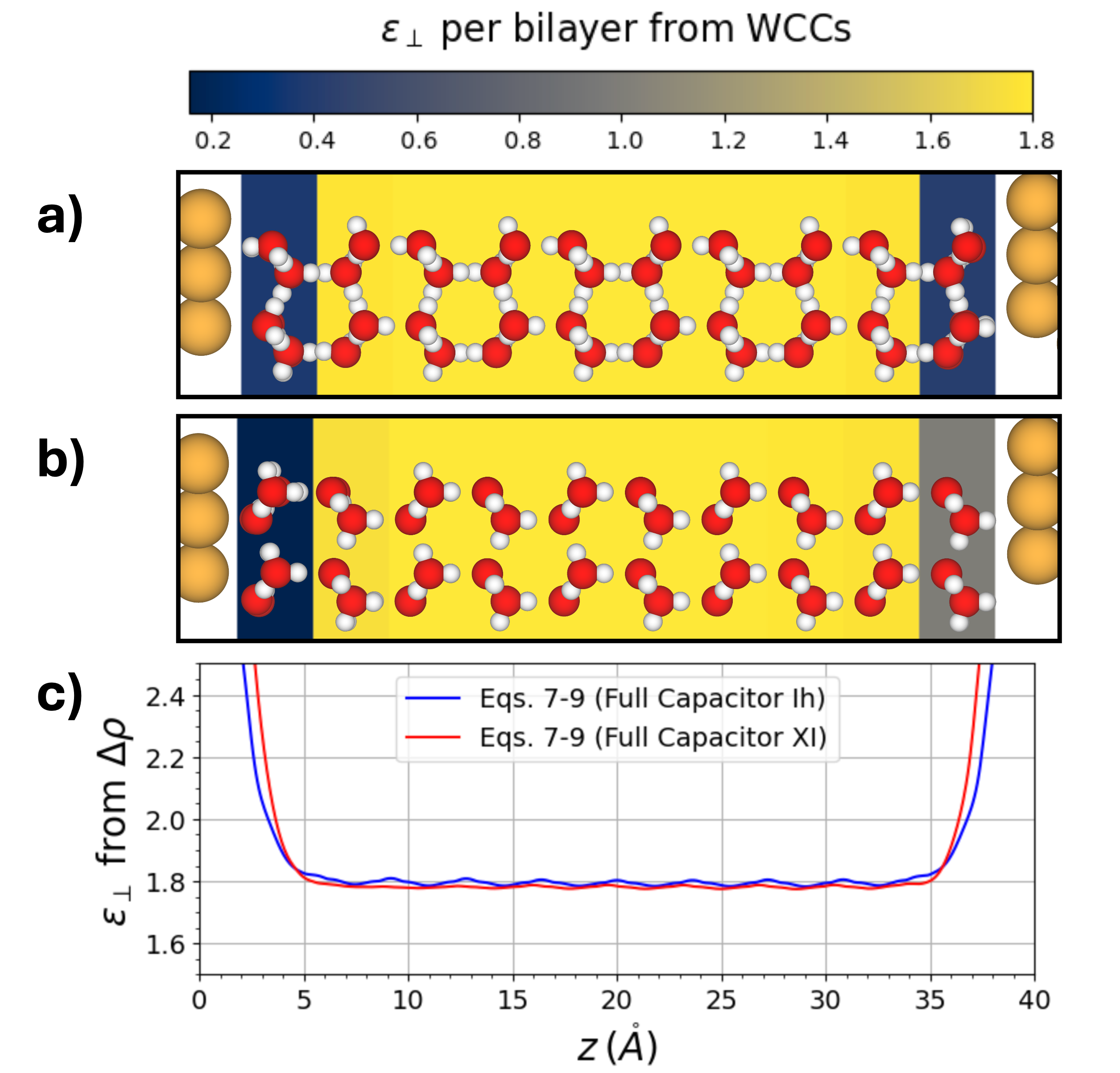}
    \caption{Optical dielectric constant $\varepsilon_{\perp}$ in full \textbf{(a)} ice \textit{Ih} and \textbf{(b)} ice \textit{XI}  capacitor geometries calculated using two methods. \textbf{(a,b)} The Wannier charge center (WCC) method described by Eqs. \ref{eq:eps} and \ref{eq:alpha_Wannier}. \textbf{(c)} The electron density method described by Eqs. \ref{eq:Gauss}-\ref{eq:ep_from P}. Surface Au electrode planes are at $z = 0$ and $z = 40$ \AA . In both cases finite differences are computed with $\Delta V=1V$.}
    \label{fig:Full_Caps}
\end{figure}

Additional care needs  to be taken when computing $\varepsilon_{\perp}$ in the ice $\it{XI}$ slab.
This is because this system is metallic at the sizes we consider in this study (5 and 10 bilayers).
This stems from inadequate screening of the ferroelectric at the ice/vacuum interface, which causes an electronic reconstruction known as the ``polar catastrophe'' \cite{nakagawa2006some}.
Nonetheless we can still compute $\Delta D_{\perp}$ for the full capacitors (see SM). Results are shown in Fig. \ref{fig:Full_Caps}.
We find that $\varepsilon_{\perp}$ within the bulk region is virtually the same for the two different proton ordered ices ($\varepsilon_{\perp}\approx1.8$).
We also see a similar suppression of the dielectric constant calculated using WCCs near the interfaces, although there seems to be a dependence on the water molecule orientation relative to the metal surface. 
A more detailed analysis of the errors in the WCCs and their dependence on the molecular orientation relative to the metal surface can be found in the SM.
We find that $w-w_0=-0.91$ {\AA} for the full ice \textit{Ih} capacitor.
%
Using Eq. \ref{eq:alpha}, we can compute the 2D polarizability with and without this width correction.
We can then compute the effective dielectric constant and the optical dielectric constant of ice using Eq. \ref{eq:eps}.
These results are shown in Table \ref{tab:table1} along with similar calculations for the standalone ice slab. 
The slab results are in noticeably better agreement with the $w\neq w_0$ result, demonstrating the importance of calculating a physically meaningful width.
Indeed, here the error due to using the incorrect width is only $5\%$ for a 10 bilayer slab of net width $\sim$ 36 \AA.  
However for a slab of net width $\sim$ 1 nm the error would increase to $20\%$.

\begin{table}[]
\centering
\begin{ruledtabular}
    \begin{tabular}{c|cccc}
    
            & $\alpha_\perp (\mathring{A})$ & $\varepsilon_\perp^{eff}$ (Eq .~\ref{eq:eps_eff})   & $\varepsilon_\perp$ (Eq. ~\ref{eq:eps})  \\ [0.5ex]
            \hline
    Au-Ice ($w=w_0$)     & 16.8 (Eq.~\ref{eq:alpha})                         & 1.80                   & 1.87          \\
    \hline
    Au-Ice ($w\neq w_0$) & 15.9 (Eq.~\ref{eq:alpha})                          & 1.76                   & 1.78            \\
    \hline
    Ice slab        & 15.7  (Eq.~\ref{eq:alpha_Wannier})                        & -                    & 1.77               \\
    \hline
    Exp.\cite{petrenko1999physics} & -& - &1.7-1.8\\
    \end{tabular}
\end{ruledtabular}    
\caption{Calculations of the 2D polarizability $(\alpha_\perp)$, effective optical dielectric constant ($\varepsilon_\perp^{eff}$), and intrinsic optical dielectric constant of ice $Ih$ ($\varepsilon_\perp$) for the full ice capacitor with $w=w_0$ and $w\neq w_0$, and the standalone ice slab.}\label{tab:table1}

\end{table}

Our results show that nanoconfinement has a negligible effect on $\varepsilon_{\perp}$ of ice, regardless of proton order. 
This is true for both a standalone ice slab and an ice capacitor with Au electrodes. 
Therefore, neither the nature of the electrodes nor confinement affect the measured 
{\it electronic} dielectric response.
According to Fumagalli {\it et al.} \cite{fumagalli2018}, this behavior is expected and underlines the anomalously low value of $\varepsilon=2.1$ measured  for the electrically dead layer of a water dielectric thin film near the capacitor interface.
However, recent studies \cite{Zubeltzu25, zhu2025dielectric} have suggested that there is a noticeably different behavior at this interface, whether it be a smaller \cite{Zubeltzu25} or larger \cite{zhu2025dielectric} electronic response.
Zubeltzu {\it et al.} \cite{Zubeltzu25} predict $\varepsilon_\perp=1.24$ for a standalone sub-nm thin film of water.
%
%
This result leads to conclude that the reduction of the total dielectric constant seen in the work of Fumagalli {\it et al.} \cite{fumagalli2018} can be mostly explained by the reduction in the electronic response. 
%
%
Zhu {\it et al.} \cite{zhu2025dielectric} use similar methods to the ones we use in this study, but come to a different conclusion. 
They assume that the electrode's charge spilling is unchanged by the presence of a dielectric, i.e. $w=w_0$, and instead attribute the moderate rise ($<25\%$) in the optical dielectric constant near the metal/water interface, calculated using the electron density, to the interfacial water molecules. 
This assignment uses the WCCs of the interfacial water molecules, which are found to be more diffuse and polarizable than their counterparts in the bulk region.
Conversely, here we have found that the WCCs are unreliable near the interface if computed from a calculation where the electrode is not decoupled.
This effect is exacerbated by the change in the capacitor geometry ($w\ne w_0$) when a dielectric is inserted.

Those last two points are important to emphasize. First, we find that the WCCs are not able to adequately partition the charge between the electrode and the dielectric, leading to specious results near the interface. 
This is not surprising in situations like this where a very small amount of charge transfer \cite{pedroza2018bias, arvelos2024probing} between the ice film and the metal takes place.
The WCCs, by construction, assume that no such charge transfer occurs.
This adds an extra level of care that needs to be taken in these interfacial calculations because not only is the width ill-defined, but the 2D polarizability can also be ill-defined if it relies on the computation of Wannier charge centers.
Our results show that this effect extends up to the third bilayer from the interface ($\sim 10$ {\AA}), which is larger than the dead layer thickness of 7.5 {\AA} reported experimentally \cite{fumagalli2018}. 
Second, we find that the electrode charge spills out further in the presence of a dielectric (i.e. $w<w_0$).
This has significant implications on Eq. \ref{eq:alpha}, which was originally derived as a way to calculate the 2D polarizability using only experimentally measurable quantities such as the capacitance \cite{Zubeltzu25}. 
The difference $w-w_0$ was intended to be the change in the distance between the atomic planes of the electrodes.
However, in our case this distance is fixed, yet we still see $w\neq w_0$. 
This means that Eq.~\ref{eq:alpha} depends on an ill-defined width, and is no longer easy to measure experimentally.

This work offers a transferable strategy for interpreting experimental capacitance data in low-dimensional dielectrics, where the standard macroscopic relation (Eq.~\ref{eq:eps_macro}) becomes only an approximate description.
In this regime, we show that differential-capacitance analyses must explicitly include electrode-induced charge smoothing via Eq.~\ref{eq:eps_eff}.
Beyond ice, this methodology can be directly applied to other 2D or nanoconfined materials—organic, hybrid, or oxide dielectrics—providing a rigorous route to their high‐frequency dielectric characterization under realistic bias. Future extensions to include ionic contributions will further bridge towards understanding the full dielectric response in nano‐scale capacitors.

{\it Acknowledgments} - MVFS is funded by the National Science Foundation award DMR-2427902 . AM thanks the Institute for Advanced Computational Science (IACS) at Stony Brook University (SBU) for a Graduate Fellowship. 
The authors would like to thank Stony Brook Research Computing and Cyberinfrastructure and IACS at SBU for access to the high-performance SeaWulf computing system, funded by the National Science Foundation (awards 1531492 and 2215987) and matching funds from the the Empire State Development’s Division of Science, Technology and Innovation (NYSTAR) program (contract C210148).
LSP, GA and ARR would like to acknowledge financial support from Funda\c{c}\~{a}o de Amparo \`a Pesquisa do Estado de S\~{a}o Paulo (FAPESP Grant \# 2017/10292-0, 2020/16593-4, 2023/09820-2), Conselho Nacional de Pesquisa (CNPq) and CAPES. Some Calculations were carried out at CENAPAD-SP and at 
the Santos Dumont High performance facilities of Laboratorio Nacional de Computa\c{c}\~{a}o Cient\'{\i}fica (LNCC), Brazil.
  This project was also partially supported by the European 
Commission Horizon MSCA-SE Project MAMBA (Grant 101131245),
the Spanish MCIN/AEI/10.13039/501100011033 through grant 
PID2022-139776NB-C65 and a Mar\'{\i}a de 
Maeztu award to Nanogune (Grant CEX2020-001038-M), 
the United Kingdom's EPSRC through Grant EP/V062654/1, 
and the Basque Ikur HPC program, through a 2023 
Cotutelage grant.

Data supporting the findings of this article are openly available \cite{github_repo}.

\bibliography{refs}
\bibliographystyle{apsrev4-2}
\appendix
\clearpage
\section*{Supplementary Material}
\addcontentsline{toc}{section}{Supplementary Material}

\section{Additional Computational Details}

\begin{figure}[!]
    \centering
     \includegraphics[width=1\linewidth]{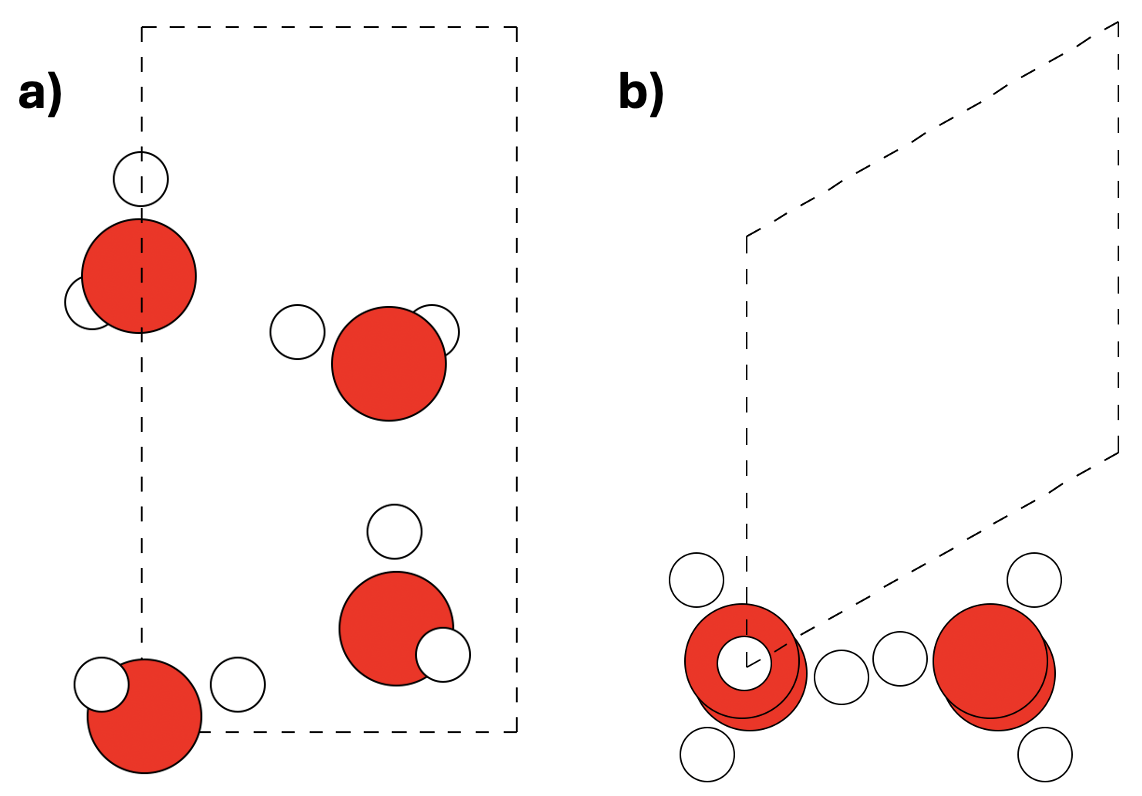}
    \caption{Illustrations of the ferroelectric unit cell used to generate all ice structures used in this work: \textbf{a)} \textit{xz}-plane and \textbf{b)} \textit{xy}-plane.}
    \label{fig:UnitCell}
\end{figure}

The original hexagonal unit cell for ferroelectric ice is shown in Fig. \ref{fig:UnitCell}.
This unit cell has four water molecules with lattice vectors $a[0,1,0]$, $\frac{a}{2}[\sqrt{3}, 1, 0]$, and $c[0,0,1]$.
The lattice parameters are $a = 4.44$ {\AA} and $c = 7.23$ {\AA}.
A $2\times 2\times 1$ supercell of this unit cell gives two bilayers of our ice \textit{XI} system.
The dipole of this supercell is then minimized using a Monte Carlo technique\cite{Buch1998} to obtain two bilayers of our ice \textit{Ih} system.
In hexagonal ice $Ih$, oxygens occupy the hexagonal wurtzite
lattice sites.
The two covalently-bonded protons have six possible orientations, but are constrained by Bernal-Fowler “ice
rules” to have one proton per tetrahedral O-O bond.
We can repeat these structures in the $c$-direction to obtain larger structures for either phase.

For the Au surface, we find that a $3 \times 3$ Au[111] surface with $a = 4.186$ {\AA} is commensurate with this ice supercell in the two in-plane directions. This lattice constant is $2.648\%$ larger than the experimental value of $a = 4.078$ {\AA}\cite{Jette1935}.
The number of Au layers used
ensures that the charge density at both edges of the simulation
box is the same as the one deep inside the bulk electrodes.
The scattering region includes three layers of the
gold electrodes.
These are then attached to six layers on each side forming the full simulation cell.
The left and right metallic surfaces are separated by $\sim$ 40 {\AA}.
This is enough space to accommodate 5 unit cells of ice, with a small gap on each side. 

\begin{figure}[!]
    \centering
     \includegraphics[width=1\linewidth]{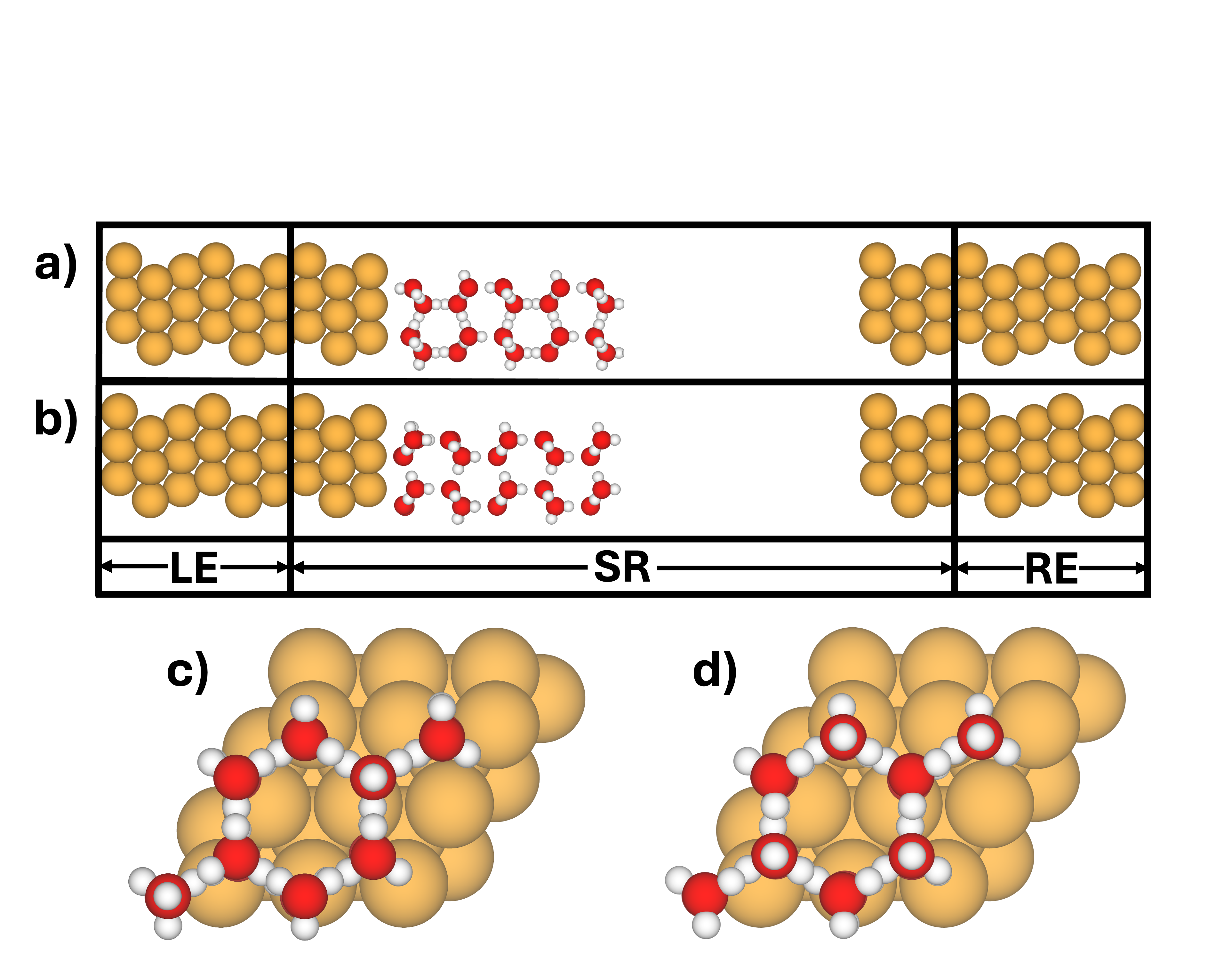}
    \caption{Illustrations of the $\mathbf{(a, c)}$ half-full ice $\it{Ih}$ - Au[111] capacitor and $\mathbf{(b, d)}$ half-full ice $\it{XI}$ - Au[111] capacitor. $\mathbf{(a, b)}$ are side views and $\mathbf{(c, d)}$ are top views. The left and right electrodes (LE and RE, respectively) and the scattering region (SR) are shown.}
    \label{fig:Half_Ice_Structs}
\end{figure}

The optimal separation distance, measured from the nearest oxygen atom, between the ice and Au surfaces was found by calculating the total energy for different separation distances and selecting the separation distance with the lowest total energy. 
This gives a separation distance of 3.3 {\AA} for ice \textit{Ih}, 2.8 {\AA} for ice \textit{XI} (O-interface), and 3.4 {\AA} for ice \textit{XI} (H-interface).
A structural relaxation was performed for the water molecules, while the Au atoms were kept fixed, until the maximum force fell below 0.01eV/{\AA}.
The 5 rightmost bilayers were then removed from the system to create the half-full capacitors, as illustrated in Fig. \ref{fig:Half_Ice_Structs}.

For the standalone ice slabs, we keep the same atomic positions and unit cell as the corresponding capacitor geometry, but we remove all of the Au atoms.
We then perform a DFT calculation without NEGF electrodes and with an external electric field applied.

All DFT calculations were carried out using the SIESTA code \cite{soler2002siesta,garcia2020siesta} within the PBE generalized-gradient approximation for exchange and correlation \cite{PBE2}. For the Au electrodes we used a custom double-$\zeta$ polarized basis set, and a custom triple-$\zeta$ basis for water's O and H atoms, all the details of these basis are included in the input files provided. We used norm-conserving Troullier–Martins pseudopotentials, a real-space mesh cutoff of 500 Ry, and $\Gamma$-point sampling of the Brillouin zone.

In order to account for the effect of the bias and consider an open system we employed the Non-equilibrium Green's function method. As discussed in the main manuscript, the system is divided into a scattering region and two semi-infinite electrodes. The key quantity is the retarded Green's function ($G^R$) of the scattering region, which includes the Kohn-Sham Hamiltonian and the so-called self-energies of the electrodes. The effects of the bias are included in the problem by shifting the chemical potential of the electrodes by $\pm \Delta V /2$ and by introducing a potential ramp in the Hamiltonian of the scattering region. From the $G^R$, it is possible to calculate the lesser Green's function, which in turn yields the non-equilibrium charge density. This establishes a self-consistent cycle analogous to standard DFT calculations. From an implementation point of view, the calculation of the NE charge density can be split into two terms: an equilibrium part and a non-equilibrium correction.
More details on the implementation of the NEGF method can be found in \cite{brandbyge2002density, pedroza2018bias, arvelos2024probing}.

For the NEGF+DFT part of the simulation, we used the same basis set, pseudopotentials, mesh cutoff, and exchange and correlation functions as in the standard SIESTA calculation. In the equilibrium part of the charge density, we used 254 Matsubara frequencies, and a total of 130 points for the complex counter (120 points for the semi-circle and 10 points for the straight line close to poles). The non-equilibrium correction, which is calculated by performing an integration along the real energy axis, employed an energy spacing of 5 meV (for V= 1 Volt this corresponds to 250 points).

All input files used to generate the calculations presented  in this paper (both for Siesta and Transiesta calculations) are openly available \cite{github_repo}.

\section{The Role of Relaxations}

As mentioned in the previous section, only the water molecules in our capacitors are relaxed, and this relaxation occurs at zero bias.
Therefore, it is important to check if our results are insensitive to (i) relaxations of the Au atoms and (ii) relaxations of the water molecules with a nonzero bias.

\begin{figure}[!]
    \centering
     \includegraphics[width=1\linewidth]{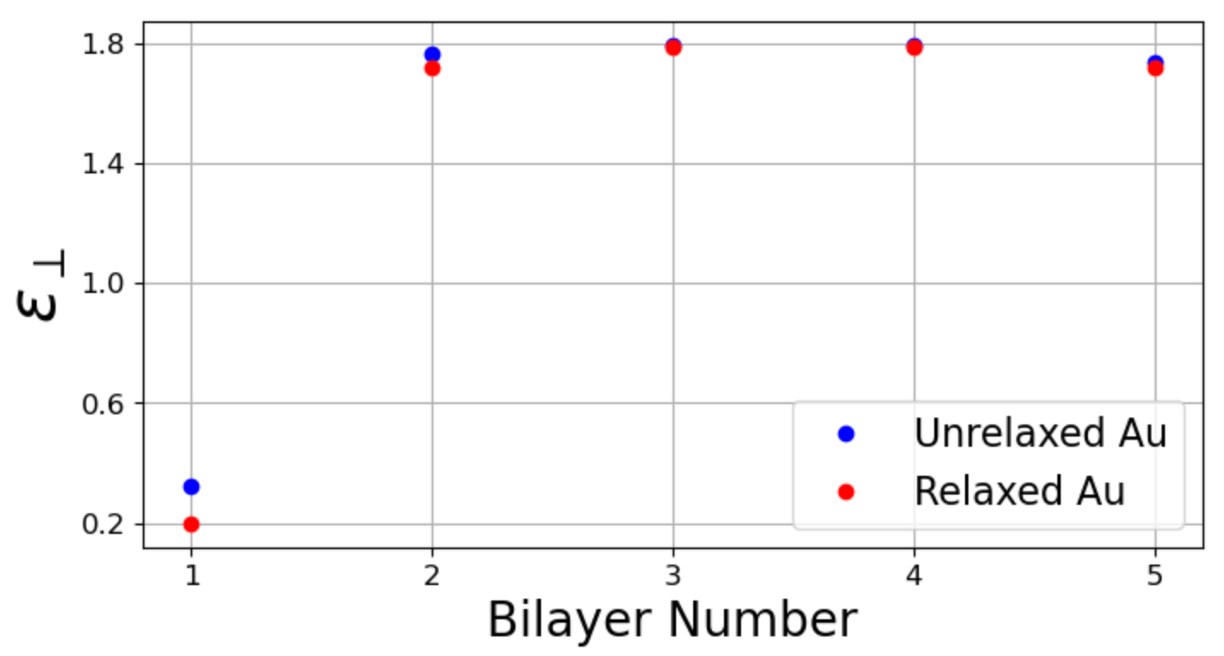}
    \caption{Optical dielectric constant, $\varepsilon_\perp$, for the half-full ice \textit{Ih} capacitor with unrelaxed (blue) and relaxed (red) Au atoms in the scattering region calculated using the WCC method.}
    \label{fig:RelaxAu}
\end{figure}

First, we performed a structural relaxation for all atoms in the scatttering region, which includes the water molecules and the first three layers of Au on either side, for our half-full ice \textit{Ih} capacitor system.
The optical dielectric constant obtained using Wannier charge centers, $\varepsilon_\perp$, is plotted in Fig. \ref{fig:RelaxAu}.
We see that relaxing the Au has a minimal effect far away from the metal-ice interface and a small, but noticeable effect near the interface.
We have already stated that the Wannier charge centers are not trustworthy near the interface, because of the difficulty in partitioning the overlapping charge between the ice bilayer and the nearby electrode.
In the new relaxed structure, the ice-metal separation distance decreases, creating more charge overlap and making it even more difficult to partition the charge.
This leads us to conclude that the small decrease in the optical dielectric constant comes from increased charge overlap between the ice and the metal, exacerbating the error in the Wannier charge centers.

\begin{figure}[!]
    \centering
     \includegraphics[width=1\linewidth]{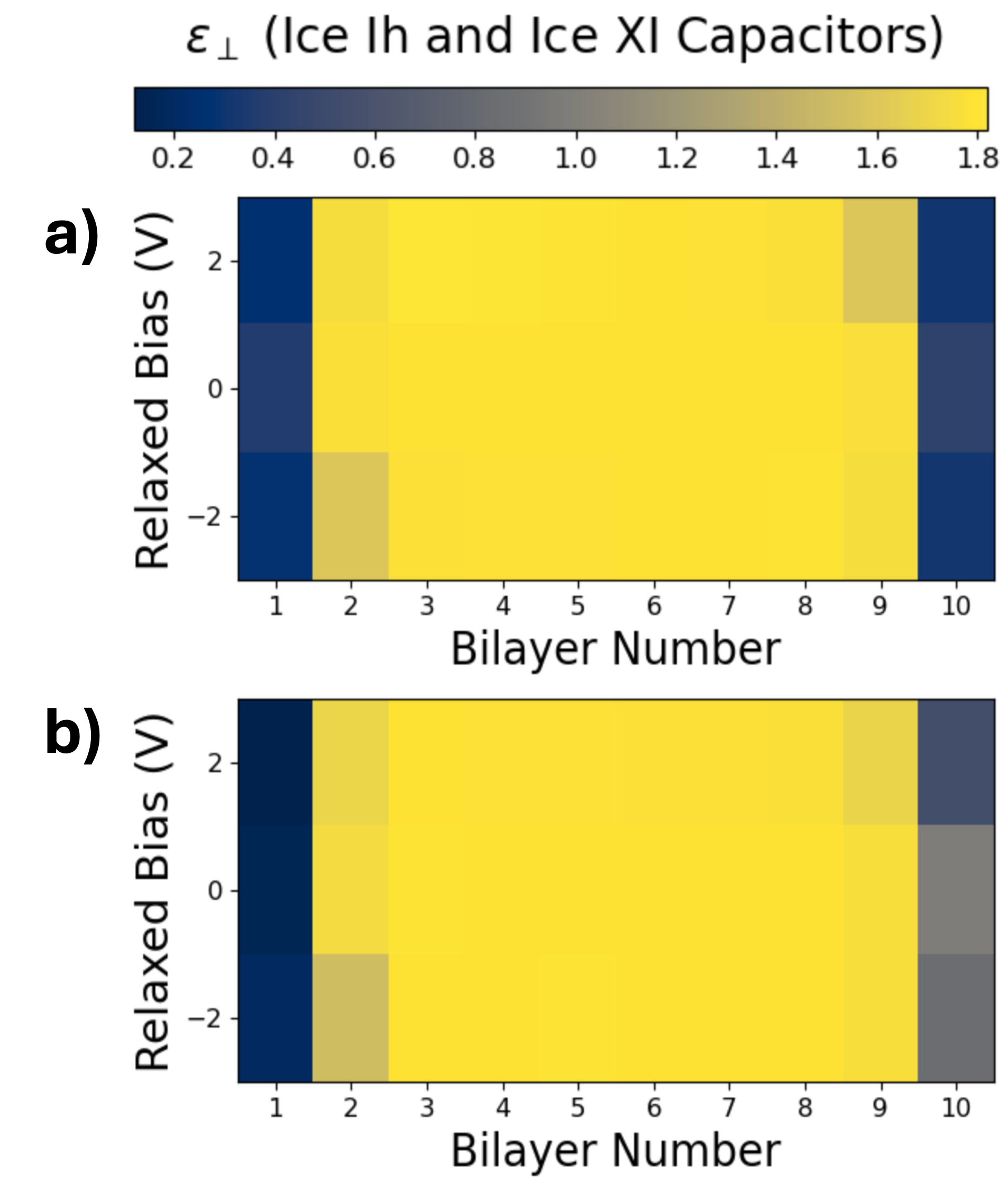}
    \caption{Optical dielectric constant, $\varepsilon_\perp$, for \textbf{(a)} the full ice \textit{Ih} capacitor and \textbf{(b)} the full ice \textit{XI} capacitor, using the WCC method. The results for the zero bias relaxed capacitor geometry are identical to the results shown in the main text. Biases of $\pm$ 2V are applied to obtain the other two relaxed capacitor geometries.}
    \label{fig:RelaxEps}
\end{figure}

Next, we performed new structural relaxations with bias in which only the water molecules were able to relax.
The optical dielectric constants calculated with different relaxed geometries for the full ice \textit{Ih} and ice \textit{XI} capacitors are shown in Fig. \ref{fig:RelaxEps}.
In the bulk region (bilayers 3-8), the optical dielectric constant remains the same (with a standard deviation less than 0.01).
There is some variance near the interface, but we attribute that to the varying charge overlap between the metal and the ice.

\begin{figure}[!]
    \centering
     \includegraphics[width=1\linewidth]{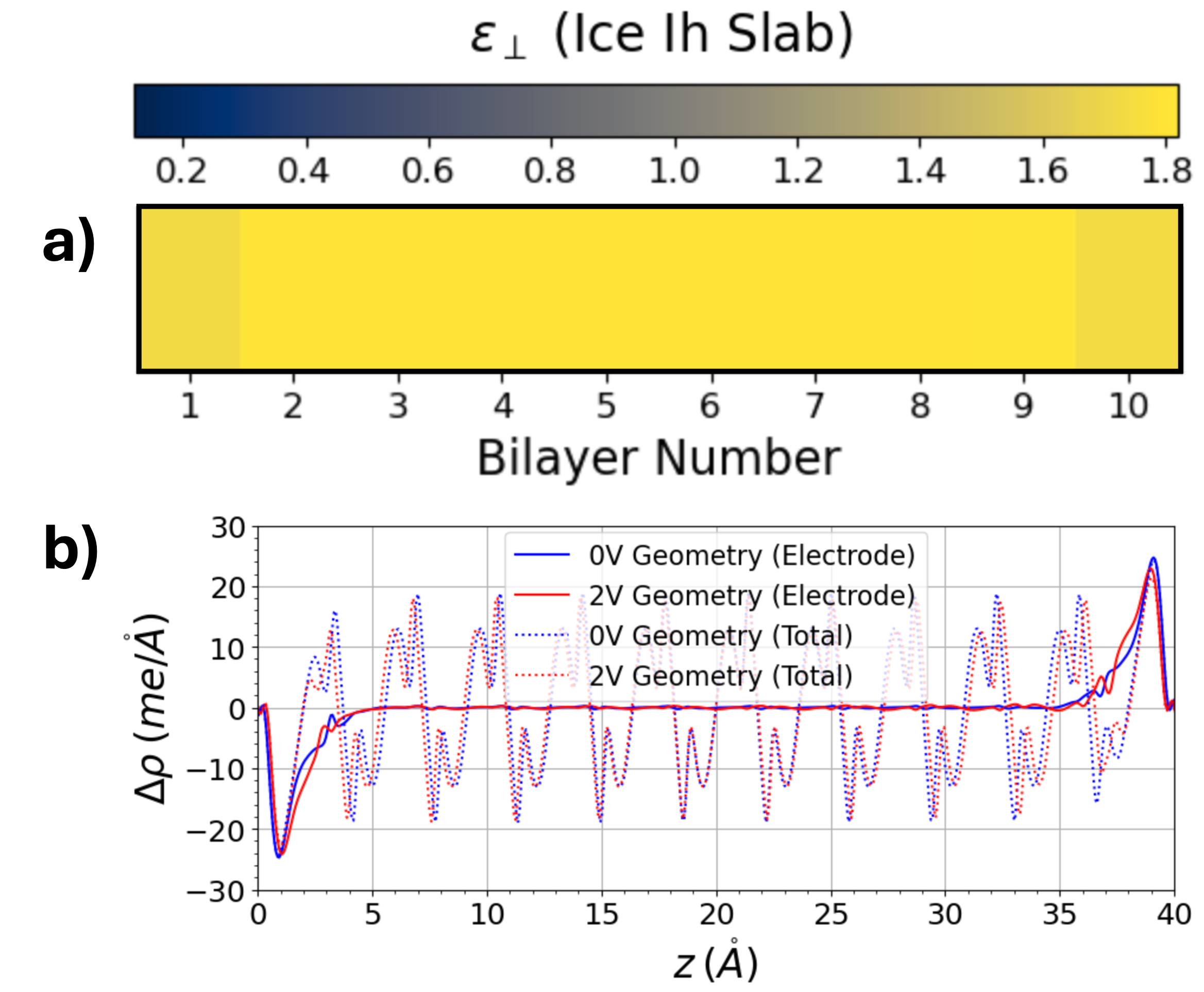}
    \caption{\textbf{a)} Optical dielectric constant, $\varepsilon_\perp$, for the ice \textit{Ih} slab, using the WCC method. \textbf{b)} Electrode charge density difference for the full ice \textit{Ih} capacitor with $V=0V$ (solid blue) and the full ice \textit{Ih} capacitor with $V=2V$ (solid red). The total charge density differences for both relaxed capacitor geometry are also shown (dotted). Surface Au electrode planes are at $z = 0$ and $z = 40$ \AA.}
    \label{fig:RelaxQ}
\end{figure}

We can also perform the charge partition between the metal and ice by subtracting the charge density of an identical ice slab from the charge density of the full capacitor.
This is shown in Fig. \ref{fig:RelaxQ}, along with the optical dielectric constant per bilayer for the ice \textit{Ih} slab.
We see that this subtraction has removed virtually all oscillations from the dielectric region, leading us to conclude that the optical dielectric response is identical to that of the ice \textit{Ih} slab.
The optical dielectric constant is 1.80 in the bulk (bilayers 2-9) and 1.73 at the surface (bilayers 1 and 10).

\section{Linearity Under Bias}

\begin{figure}[!]
    \centering
     \includegraphics[width=1\linewidth]{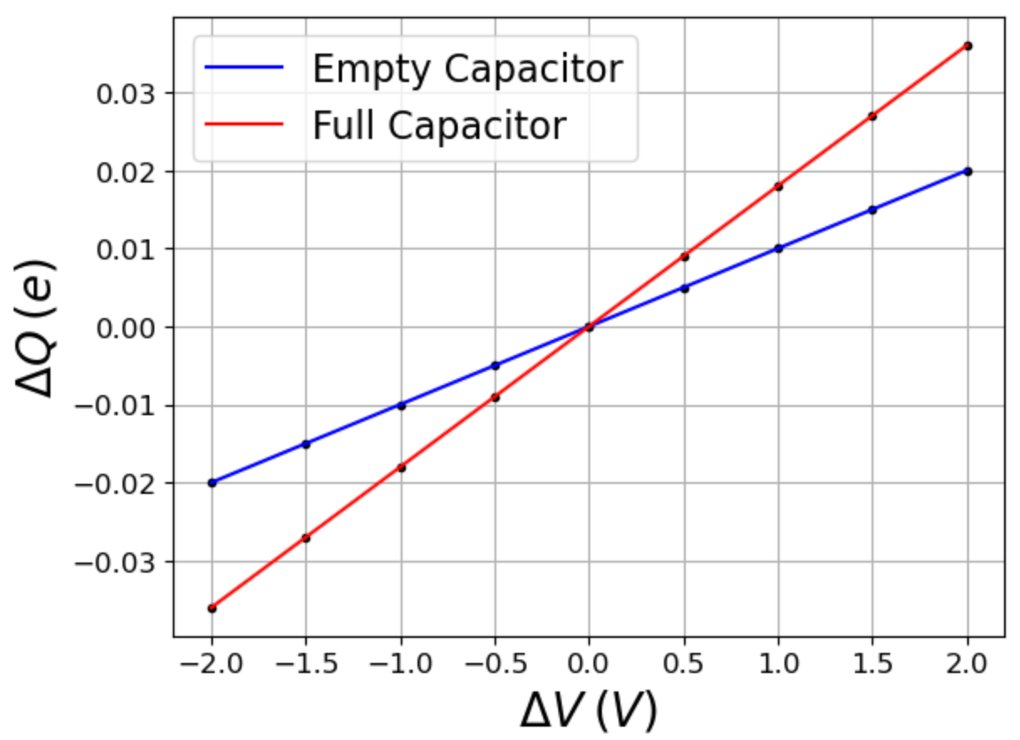}
    \caption{Electrode charge difference, $\Delta Q$ as a function of the applied bias difference $\Delta V$ for the empty capacitor and the full ice \textit{Ih} capacitor.}
    \label{fig:QvV}
\end{figure}

It is also important to check that our results are linear with respect to the applied bias.
This means that the electrode charge should increase linearly with the applied voltage, with the slope being the capacitance.
We have evaluated the electrode charge difference for the empty capacitor and the full ice \textit{Ih} capacitor (using our charge partition method).
These results are plotted in Fig. \ref{fig:QvV}.
Both plots are linear, and the ratio between the two slopes gives 1.80. Multiplying this by $w/w_0$, we get the effective optical dielectric constant $\varepsilon_\perp^{eff}=1.76$.

We can also evaluate the linearity under bias by evaluating the 2D polarizability, $\alpha_\perp$, and the optical dielectric constant, $\varepsilon_\perp$, for different applied bias differences, |$\Delta V$|.
These results are shown in Table \ref{tab:table1_SI}.
They show that both quantities are insensitive to $\Delta V$, especially in the bulk region.
It should be noted that the 2D polarizability and dielectric constant at the surface, while relatively constant under bias, are inaccurate due to the errors in the WCCs when the dielectric and electrode charge overlap.

We take advantage of this linearity under bias when performing the standalone slab calculation.
Instead of evaluating this system for a displacement field difference $\Delta D_\perp^{Slab}$ exactly equal to that of the capacitor $\Delta D_\perp^{Cap.}$, we can instead evaluate it for an arbitrarily chosen value of $\Delta D_\perp^{Slab}$ and then evaluate the charge density difference in the standalone slab geometry $\Delta\rho^{Slab}(z)$ for a given $\Delta D_\perp^{Cap.}$ in the following way:

\beq
\label{eq:convert_D}
\Delta \rho^{Slab}\big(z,\:\Delta D_\perp^{Cap.}\big) = \Delta \rho^{Slab}\big(z,\:\Delta D_\perp^{Slab}\big)\cdot\frac{\Delta D_\perp^{Cap.}}{\Delta D_\perp^{Slab}}.
\eeq

This is useful because if we were unsure of the displacement field in the capacitor, we could arbitrarily choose a value of $\Delta D_\perp^{Slab}$, perform a standalone slab calculation to determine $\Delta \rho^{Slab}\big(z,\:\Delta D_\perp^{Cap.}\big)$ for different values of $D_\perp^{Cap.}$, and subtract these results from $\Delta\rho^{Cap.}(z)$ to see at what value of $D_\perp^{Cap.}$ the oscillations in the center of the dielectric region are suppressed.
This value of $D_\perp^{Cap.}$ is the actual displacement field difference of the capacitor.

\begin{table}[]
\centering
\begin{ruledtabular}
    \begin{tabular}{c|cccc}
    
            $|\Delta V| (V)$
            & $\alpha_\perp^{Bulk} (\mathring{A})$
            & $\varepsilon_\perp^{Bulk}$
            & $\alpha_\perp^{Surface} (\mathring{A})$
            & $\varepsilon_\perp^{Surface}$  \\ [0.5ex]
            \hline
    0.5  & 1.602 $\pm$ 0.001                         & 1.795 $\pm$ 0.001                   & -5.4 $\pm$ 0.7  & 0.41 $\pm$ 0.03       \\
    \hline
    1.0 & 1.599 $\pm$ 0.001                         & 1.794 $\pm$ 0.001                   & -4.7 $\pm$ 0.6 & 0.44 $\pm$ 0.03           \\
    \hline
    1.5 & 1.598 $\pm$ 0.002  & 1.792 $\pm$ 0.001                    & -4.6 $\pm$ 0.7  & 0.44 $\pm$ 0.04             \\
    \hline
    2.0 & 1.597 $\pm$ 0.002 & 1.791 $\pm$ 0.002 & -6.2 $\pm$ 0.8 & 0.37 $\pm$ 0.03\\
    \end{tabular}
\end{ruledtabular}  
\caption{Calculations of the 2D polarizability $(\alpha_\perp)$ and optical dielectric constant of ice \textit{Ih} $\varepsilon_\perp$ for a single bilayer of ice \textit{Ih} in the full capacitor geometry with an applied bias difference |$\Delta V$|. Bilayers are split into bulk bilayers (bilayers 3-8) and surface bilayers (bilayers 1 and 10). The average value and standard deviation for each quantity are tabulated. The results measured at the surface are inaccurate due to errors in the WCCs stemming from charge overlap at the surface.}
\label{tab:table1_SI}
\end{table}

\section{The Correct Capacitor Width}

For the empty capacitor, we can easily compute the capacitance as $C_0=Q_0/V$, which gives us $C_0=1.602\cdot10^{-21}F$. Equivalently, we can compute the capacitance from the geometry as $C=\epsilon_0A/w_0$. Using our previously obtained capacitance and solving for $w_0$, we obtain $w_0 = 37.73$ {\AA}. This is $\sim2$ {\AA} less than the distance between the two interfacial Au atomic planes, $w_{atom} \approx 40$ {\AA}. We can also compute $w_0$ using the centroid of the charge in each electrode.

\beq
\label{eq:z_charge}
z_{charge} = \frac{\int{z\Delta\rho(z)dz}}{\int{\Delta\rho(z)dz}}
\eeq

The integration bounds are located in the regions on either side of the electrode where $\rho(z)=0$. $w_{charge}$ can be computed in the following way.

\beq
\label{eq:w_charge}
w_{charge} = z_{charge}^{right}-z_{charge}^{left}
\eeq

This gives $w_{charge}=37.69$ {\AA}, which is a much better estimate of $w_0$ than $w_{atom}.$

\section{A Naive Electron Density Partition Method}
\label{sec:e-density}

This section highlights the difficulty in partitioning the electron density between the electrode and dielectric, or between different layers of the dielectric.

In order to calculate $\alpha_{\perp}$ and $\epsilon_{\perp}$ for each bilayer, we must first compute the change in charge on the electrodes, $\Delta Q_{elec}$, and the change in 2D polarization per bilayer, $\Delta \mathcal{P}_{2D}$, as the applied bias, $V$, changes. At first glance, both quantities should be easily computed using the electron density, $e(\vec{r})$. We use the electron density to create two measures of charge differences as a function of the out-of-plane direction, $z$, and the initial and final biases, $V_1$ and $V_2$.

\beq
\label{eq:drho}
\Delta\rho(z,\{V_{1},V_{2}\}) = \bigg[- \iint e(x,y,z) dx dy \bigg]_{V_{1}}^{V_{2}}
\eeq

\beq
\label{eq:dQ}
\Delta Q(z,\{V_{1},V_{2}\}) = \bigg[\int_0^z \Delta \rho(z') dz'\bigg]_{V_{1}}^{V_{2}}
\eeq

\begin{figure}[!]
    \centering
     \includegraphics[width=1\linewidth]{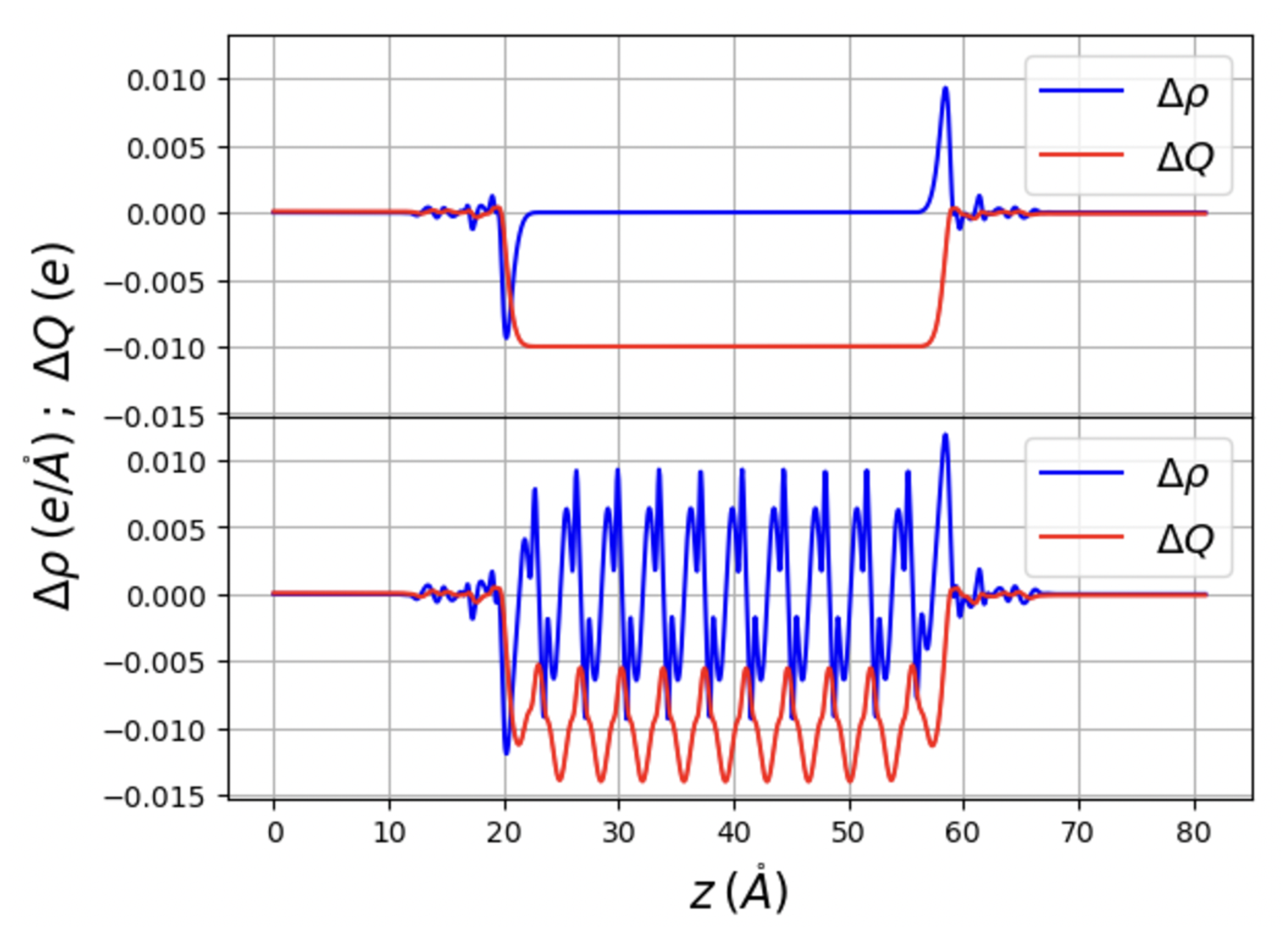}
    \caption{$\Delta \rho (z)$ and $\Delta Q (z)$ for the empty capacitor (top) and the full ice $\it{Ih}$ - Au capacitor (bottom).}
    \label{fig:Ice_Ih}
\end{figure}

\noindent Here, $\Delta \rho(z)$ is a one-dimensional charge density with units of charge per unit length (integrated over the two in-plane directions). $\Delta Q(z)$ is a cumulative integral of $\Delta \rho(z)$ and has units of charge. Both of these quantities are plotted for an empty capacitor and a full capacitor in Fig. \ref{fig:Ice_Ih}. $\Delta \mathcal{P}_{2D}$ and $\Delta Q_{elec}$ can then be calculated using the following equations:

\beq
\label{eq:dP_2D}
\Delta \mathcal{P}_{2D} = \frac{1}{A} \int_{z_1}^{z_2} z \Delta \rho(z) dz
\eeq

\beq
\label{eq:dQ_C}
\Delta Q_{elec} = \max | \Delta Q(z) |
\eeq

\noindent where $A$ is the cross-sectional area of the capacitor, and $z_1$ and $z_2$ are the bounds of the bilayer of interest. These bounds can be determined by finding the location of local maxima of $\Delta Q(z)$ shown in Fig. \ref{fig:Ice_Ih}. 

\begin{figure}[!]
    \centering
     \includegraphics[width=1\linewidth]{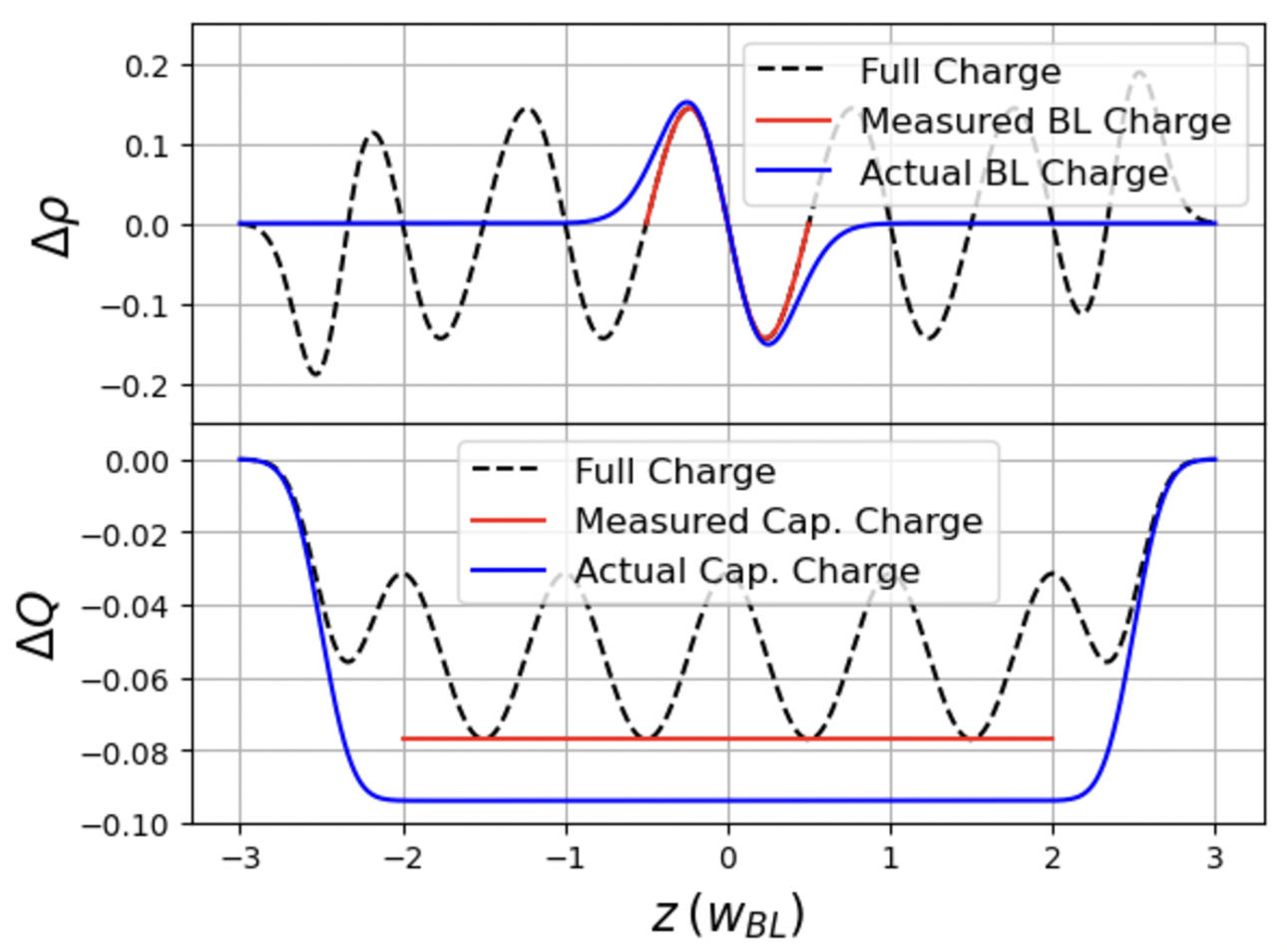}
    \caption{$\Delta \rho (z)$ (top) and $\Delta Q(z)$ (bottom) for a five bilayer capacitor where the 1D charge density of a single bilayer is given by $\Delta \rho_{BL}(z, \{\mu_{BL}, \sigma_{BL}\}) = (z-\mu)\exp\bigl\{-\frac{(z-\mu_{BL})^2}{2\sigma_{BL}^2}\bigl\}$ with $\sigma_{BL}=0.25$ and $\mu_{BL}=-2,-1,0,1,2$. This scales the horizontal axis to be in units of the thickness of one bilayer, $w_{BL}$. The surface charge from the metal is given by $\Delta \rho_{M}(z) = \mp0.25(z\pm2.5)\exp\bigl\{-\frac{(z\pm2.5)^2}{2(0.15)^2}\bigr\}$. The scales of the vertical axes are arbitrary. The black dashed line shows the charge of the full five bilayer capacitor the red line shows the suppressed measured charge, and the blue line shows the actual charge.}
    \label{fig:BL_Model}
\end{figure}

This method is attractive because of its simplicity, but it fails to properly calculate both quantities of interest. This can be illustrated using the toy model shown in Fig. \ref{fig:BL_Model}. Here, we model each bilayer as a polarized Gaussian and the surface charge from the metal as two oppositely charged Gaussians:

\beq
\label{eq:BL_Model}
\Delta \rho_{BL}(z, \{\mu_{BL}, \sigma_{BL}\}) = (z-\mu_{BL})\exp\Biggl\{-\frac{(z-\mu_{BL})^2}{2\sigma_{BL}^2}\Biggr\}
\eeq

\beq
\label{eq:M_Model}
\Delta \rho_{M}(z, \{\mu_M, \sigma_M\}) = \mp A_M\exp\Biggl\{-\frac{(z\pm\mu_M)^2}{2\sigma_M^2}\Biggr\}
\eeq

\noindent where $(\mu_{BL}, \sigma_{BL})$ and $(\mu_{M}, \sigma_{M})$ are the center position and width for each bilayer and the metal, respectively. $A_M$ is a constant that ensures a reasonable ratio between the bilayer charge and the metallic charge.

When a bilayer is placed between two other bilayers, the left (right) tail of the central bilayer is suppressed by the right (left) tail of the left (right) bilayer. This can lead to a significant underestimation of $\Delta \mathcal{P}_{2D}$ per bilayer. Similarly, the charge on the left (right) electrode is suppressed by the left (right) tail of the nearest bilayer. This can lead to a significant underestimation of $\Delta Q_{elec}$.

\section{Wannier Charge Centers (WCCs)}

We can alternatively calculate the local dipole moments of individual molecules using Wannier functions $\ket{w_{i}({r_{i}})}$, where $\bm{r_{i}}$ are the Wannier charge centers (WCCs).
The Bloch states $\ket{\psi_{n\bm{k}}}$ are obtained from a $\Gamma$ point calculation for the scattering region of the system.
These are obtained using the converged non-equilibrium Green's function (NEGF) calculation electron density in a non self-consistent periodic calculation.

The 2D polarization per bilayer, $\mathcal{P}_{2D}^{BL}$ is given by

\beq
\label{eq:P_2D}
\mathcal{P}_{2D}^{BL} = \frac{1}{A}\sum_{i,I}\big(-2ez_i+q_Iz_I\big)
\eeq

\noindent where $i$ and $I$ are indices for the WCCs and atomic coordinates, respectively. 
$A$ is the cross-sectional area of the bilayer, $e$ is the elementary charge, and $q_I$ is the nuclear+core charge for each atom.

\section{The Displacement Field}

It is important to note that we include two different DFT calculation techniques and the displacement field change, $\Delta D_{\perp}$, is computed differently for each technique.
When we apply a bias using the NEGF method, this is akin to connecting the capacitor to a battery.
This means that the displacement field change is computed in the following way:

\beq
\label{eq:D1}
\Delta D_{\perp} = -\epsilon_0\frac{d\Delta V_H}{dz} + \frac{\Delta\mathcal{P}_{2D}}{w}
\eeq

\noindent where $\Delta V_H(z)$ is the Hartree potential difference between two applied biases. We can clearly see here that it is preferable to compute $\Delta D_\perp$ in vacuum, since $\mathcal{P}_{2D}=0$ in that region and $w$ does not need to be well-defined.
However, this is not the only route to obtaining $\Delta D_\perp$. 
As we have shown, the 2D polarization change $\Delta\mathcal{P}_{2D}$ is well-defined for bilayers within the bulk region using the WCC method.
$w$ is also well-defined for a crystalline system like ice.

For a DFT calculation without NEGF electrodes where we apply an external electric field, the displacement field change is simply equal to the external electric field change.

\beq
\label{eq:D1}
\Delta D_{\perp} = \epsilon_0 \Delta E_{ext}
\eeq

Since $\Delta \rho(z)$ and $\Delta Q(z)$ are directly proportional to $\Delta D_{\perp}$, we can easily convert our results between displacement fields using the ratio between the two fields.

\section{The Polar Catastrophe in Ice XI}

\begin{figure}[!]
    \centering
     \includegraphics[width=1\linewidth]{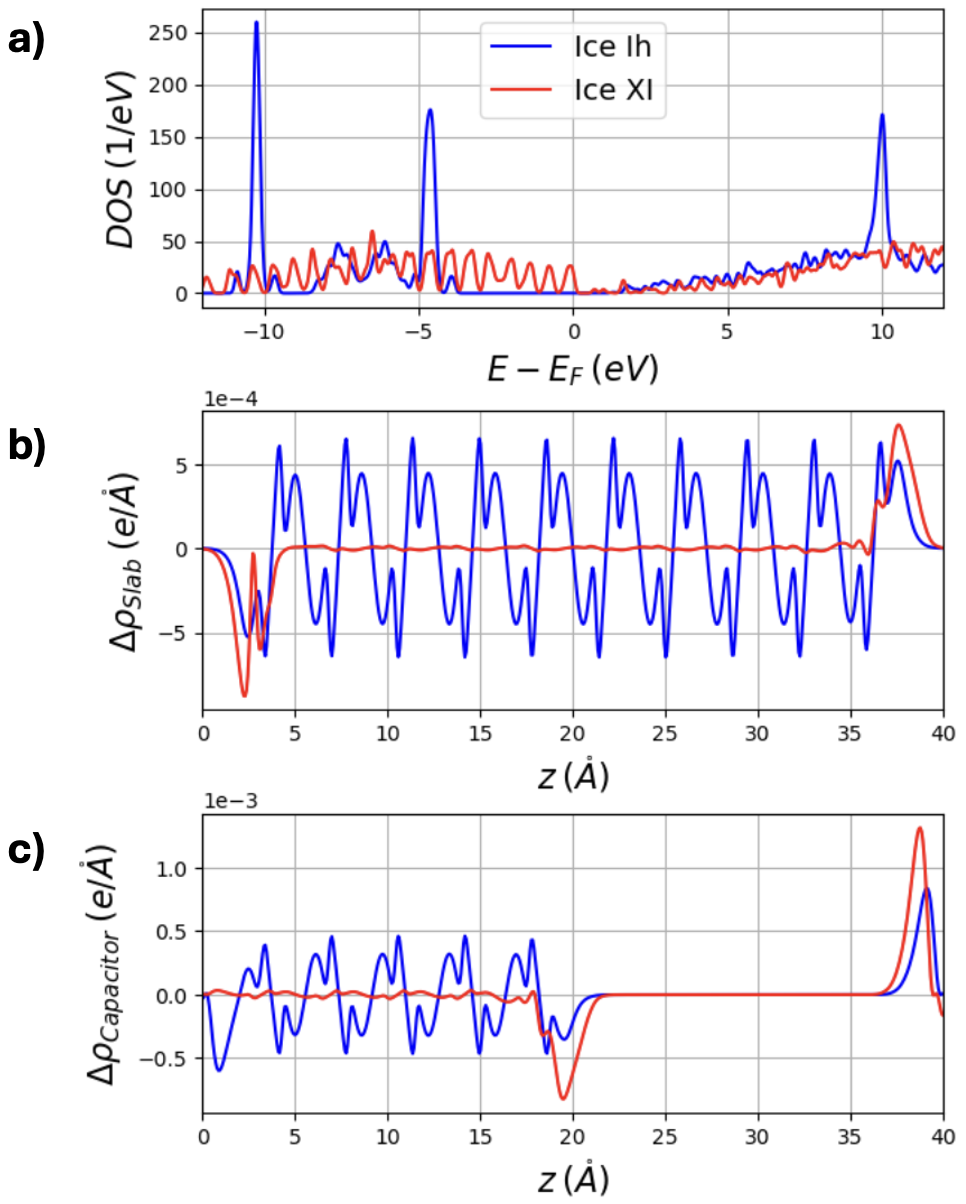}
    \caption{$\mathbf{a)}$ The density of states at zero bias for the ten bilayer standalone slabs of ice $\it{Ih}$ (blue) and ice $\it{XI}$ (red). $\Delta \rho (z)$ for $\mathbf{(b)}$ the two ice slabs and $\mathbf{(c)}$ the two half-full ice capacitors. Surface Au electrode planes are at $z = 0$ and $z = 40$ \AA.}
    \label{fig:Ice_XI}
\end{figure}

When we perform a DFT calculation of the ice \textit{XI} slab (without NEGF electrodes) and the half-filled ice \textit{XI} capacitor (with the NEGF method), we observe a strange electronic phenomenon.
As shown in Fig. \ref{fig:Ice_XI}, the ice charge is concentrated near the interfaces, similar to what one would expect for a metal.
For ice \textit{Ih}, its charge is distributed throughout the slab, as expected for a typical dielectric. Furthermore, when we plot the density of states for both systems in Fig. \ref{fig:Ice_XI}, we see that there are states at the Fermi level for ice \textit{XI}.
This suggests a metallic behavior that can be found in ferroelectric materials known as the "polar catastrophe", which stems from insufficient screening of the ice/vacuum interface.

\section{Molecular Orientation and WCC Error}

\begin{figure}[!]
    \centering
     \includegraphics[width=1\linewidth]{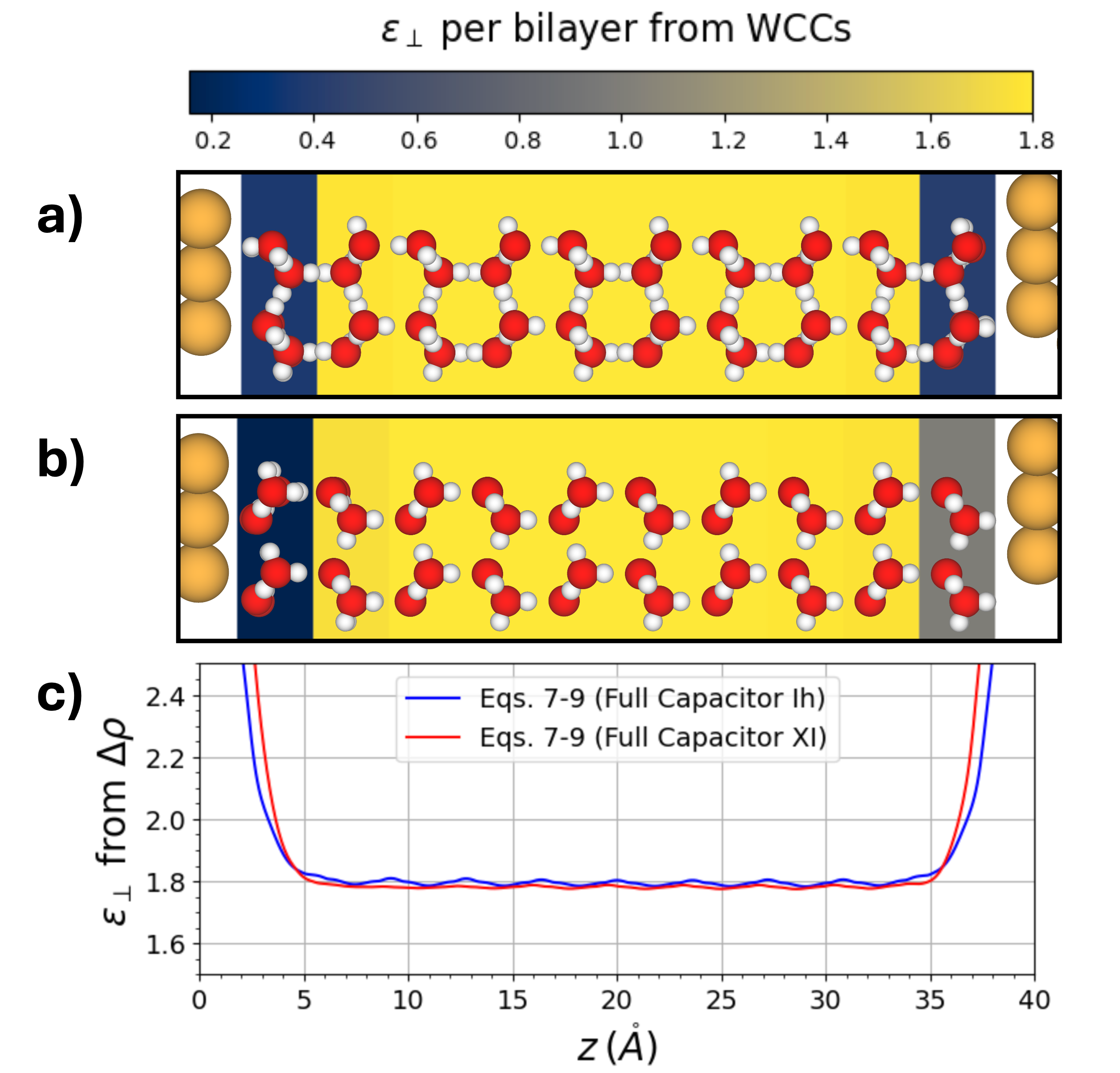}
    \caption{Optical dielectric constant $\varepsilon_{\perp}$ in full \textbf{(a)} ice \textit{Ih} and \textbf{(b)} ice \textit{XI}  capacitor geometries calculated using two methods. \textbf{(a,b)} The Wannier charge center (WCC) method. \textbf{(c)} The electron density method. Surface Au electrode planes are at $z = 0$ and $z = 40$ \AA . In both cases finite differences are computed with $\Delta V=1V$.}
    \label{fig:Full_Caps_SI}
\end{figure}

In Fig. \ref{fig:Full_Caps_SI} (shown in the main text and replicated here), we show the optical dielectric constant of both ice capacitors for each of the 10 bilayers, computed using Wannier charge centers. We conclude that near the ice/metal interface these Wannier charge centers are ill-defined, due to the difficulty in accurately partitioning the charge between the ice and the metal.

We can gain more insight into this error by looking into the magnitude of the error in different situations. For example, in ice \textit{Ih} the two ice/metal interfaces are symmetric. This means that for the four water molecules in contact with the Au electrode on either surface, two have oxygen atoms facing the metal and two have hydrogen atoms facing the metal. In Fig. \ref{fig:Full_Caps_SI}, we see that the optical dielectric constants computed for both of these interfacial bilayers are nearly identical, which is consistent with this symmetry. For ice \textit{XI}, the interfaces are antisymmetric. The leftmost bilayer has four oxygen atoms facing the metal and the rightmost bilayer has four hydrogen atoms facing the metal, and this antisymmetry is reflected in the calculated optical dielectric constants. We can see that the error from the Wannier charge centers is much more prevalent in the leftmost bilayer, where the oxygen atoms face the metal. This makes sense because when the oxygen atoms face the metal, the lone pairs from each water molecule overlap with the surface charge of the electrode, making it even more difficult to discern between the electrode charge and the lone pairs from the water molecules.


\end{document}